\documentstyle[prd,aps,eqsecnum,epsfig]{revtex}

\begin{document}


\hspace{3.7in}February 18, 1998

\vspace{2.0cm}

\noindent 
Mr. D. L. Nordstrom\\
Editor - Physical Review D
1 Research Road, Ridge, New York

\vspace{0.25in}

\noindent
Dear Mr. Nordstrom

\noindent
We are submitting our manuscript entitled {\it Chaos in preinflationary Friedmann-Robertson-Walker universes} for publication in Physical Review D as a regular article.

\noindent
Sincerely,

\vspace{1.6cm}

Henrique P. de Oliveira

\newpage

\title{Chaos in Preinflationary Friedmann-Robertson-Walker Universes}

\author{G. A. Monerat\thanks{E-mail: germano@symbcomp.uerj.br}, H. P. de
Oliveira\thanks{E-mail: oliveira@symbcomp.uerj.br}}

\address{{\it Universidade do Estado do Rio de Janeiro }\\ 
{\it Instituto de F\'{\i}sica - Departamento de F\'{\i}sica Te\'orica,}\\ 
{\it CEP 20550-013 Rio de Janeiro, RJ, Brazil}}

\author{I. Dami\~ao Soares\thanks{E-mail: ivano@lca1.drp.cbpf.br.}}

\address{{\it Centro Brasileiro de Pesquisas F\'\i sicas\\ Rua Dr. 
Xavier Sigaud, 150 \\ CEP 22290, Rio de Janeiro~--~RJ, Brazil}}

\maketitle 

\thispagestyle{empty} 

\label{address}
\vspace{0.34cm} 
\noindent

\begin{abstract}
The dynamics of a preinflationary phase of the universe, and its exit to
inflation, is discussed. This phase is modeled by a closed
Friedmann-Robertson-Walker geometry, the matter content of which is radiation
plus a scalar field minimally coupled to the gravitational field. The
energy-momentum tensor of the scalar field is split into a cosmological
constant type term, corresponding to the vacuum energy of the scalar field
plus the energy-momentum tensor of the spatially homogeneous expectation
value of the scalar field. This simple configuration, with two effective
degrees of freedom only, presents a very complicated dynamics connected to
the existence of critical points of saddle-center-type and saddle-type in the
phase space of the system. Each of these critical points is associated to an
extremum of the scalar field potential. The topology of the phase space about
the saddle-centers is characterized by homoclinic cylinders emanating from
unstable periodic orbits, and the transversal crossing of the cylinders, due
to the non-integrability of the system, results in a chaotic dynamics. The
topology of the homoclinic cylinders provides an invariant characterization
of chaos. The model exhibits one or more exits to inflation, associated to
one or more strong asymptotic de Sitter attractors present in phase space,
but the way out from the initial singularity into any of the inflationary
exits is chaotic. We discuss possible mechanisms, connected to the spectrum
of inhomogeneous fluctuations in the models, which would allow us to
distinguish physically the several exits to inflation.

\end{abstract}

\label{PACS number:47.75+f}

\vfill\eject

\newpage

\section{Introduction}

The existence of an inflationary phase in the early stages of our Universe
has become one of the paradigms of modern cosmology\cite{inflation}, and is
now being subject to experimental verification through the crucial
measurements of small scale anisotropies in the cosmic background
radiation. The basic physical ingredient for this inflationary phase is the
existence of a scalar field - the inflaton field, the vacuum energy of which
plays the role of a cosmological constant, engendering via the gravitational
dynamics an exponential expansion in the comoving scales of the
universe. This model may be thought to have evolved from a pre-inflationary
phase just exiting the Planck era. Our attempt here will be to discuss a
model for this pre-inflationary phase and its exit to inflation, by using a
minimal set of ingredients: in its simplest version the model can be
described as a FRW universe, the matter content of which is a perfect fluid
(which we take as radiation) plus a scalar field minimally coupled to the
gravitational field. The energy momentum tensor of the scalar field may be
split into a cosmological constant-type term (corresponding to the vacuum
energy of the scalar field) plus the energy momentum tensor of the spatially
homogeneous expectation value of the scalar field. We assume from the start a
closed FRW universe. This apparently simple configuration, with two effective
degrees of freedom only, will present a very complicated dynamics. In fact,
we will show that $(i)$ the model exhibits one or more exits to inflation,
associated to one or more asymptotic de Sitter attractors in phase space,
depending on the structure of the scalar field potential; $(ii)$ the dynamics
will imply that in any of the exits to inflation the scalar field will be
frozen in one of vacuum states (with its corresponding vacuum energy playing
the role of the cosmological constant for that exit) associated to an
extremum of the scalar field potential; $(iii)$ the exit to inflation is
chaotic.

In the literature of inflation, the exponential expansion of the scales of
the model, whenever a positive cosmological constant is present, has been
extensively discussed and constitutes the basis of the so-called cosmic
no-hair conjecture. Wald\cite{wald}, in the realm of spatially homogeneous
cosmologies, and Starobinskii\cite{starob}, in the more general case of
inhomogeneous models, showed that all initially expanding models evolve
towards the de Sitter configuration if a positive cosmological constant is
present. Also, the introduction of dynamical degrees of freedom associated to
the scalar fields, in addition to a cosmological constant, was shown to
produce nontrivial dynamics in the early stages of inflation. In particular,
Calzetta and El Hasi\cite{calza}, and Cornish and Levin\cite{cornish}
exhibited chaotic behavior in the dynamics of FRW models with a cosmological
constant term and scalar fields conformally and/or minimally coupled to
gravitation, implying that small fluctuations in initial conditions of the
model preclude or induce the Universe to inflate. In both cases, the presence
of the cosmological constant induces the existence of separatrices in the
unperturbed phase space of the models, connecting critical points of the
saddle-type. These connections are known to be highly unstable, and their
breaking and transversal crossing, due to perturbations originated from the
coupling of the gravitational variable with the scalar field, are basically
responsible for the chaotic dynamics. This chaotic behavior constitutes the
so-called Poincare's homoclinic phenomena\cite{ivano}.

The new features in the preinflationary dynamics introduced by our model
arise from the presence of a positive cosmological constant and a perfect
fluid (matter in radiation form) in a closed FRW universe, whenever a scalar
field is also present even in the form of small perturbations, and have not
been considered in the literature yet. The degrees of freedom are the scale
factor of the FRW geometry and one scalar field. As a consequence, the phase
space of the system present one or more critical points, depending on the
number of extrema of the potential of the scalar field. If the extrema is a
minimum, the corresponding critical point is identified as a
saddle-center. As a consequence, we have a very complex dynamics based on the
so-called homoclinic cylinders which emanate from unstable periodic orbits
that exist in a neighborhood of a saddle-center. Analogous to the breaking
and crossing of heteroclinic curves in above referred homoclinic phenomena,
these cylinders will cross each other in the non-integrable cases producing a
chaotic dynamics. These topological structures allow an invariant
characterization of chaos in the model\cite{oss}, as we will discuss, and
will have a deep implication for the occurrence or not of inflation in its
several exits, as well as for the physics in the early stages of
inflation. Chaotic exit to inflation, as a consequence of the existence of a
saddle-center and its associated structure of homoclinic cylinders, has been
examined for the first time in the literature by de Oliveira, Soares and
Stuchi\cite{oss} (subsequently referred to OSS), and contains several
technical results which we will often refer to here. The paper is organized
as follows. In Section II, we describe the Hamiltonian dynamics of the model
and the critical points in the phase space. We discuss the nature of these
critical points, and describe the topology of homoclinic cylinders about each
saddle-center critical point. The numerical evidence for a physically
relevant chaotic behavior is showed in Sec. III, where two cases concerning
the choices of the potential will be discussed. Finally, in Sec. IV, we
conclude and trace some perspectives of the present work.

\section{The Dynamics of the Model}

We consider closed FRW cosmological models characterized by the scale
function $a(t)$, with line element

\begin{equation}
d\,s^2 = d\,t^2 - a^2(t)\,[(w^1)^2 + (w^2)^2 + (w^3)^2].                                        
\end{equation}

\noindent Here $t$ is the cosmological time and $(w^1, w^2, w^3)$ are
invariant Bianchi type-IX 1-forms\cite{mtw}. The matter content of the model
is represented by a perfect fluid with four-velocity $U^{\mu} =
\delta^{\mu}_0$ in the comoving coordinate system used, plus a homogeneous
minimally coupled scalar field with potential $V(\varphi)$. The total energy
momentum tensor is described by

\begin{equation}
T_{\mu\nu} = (\rho + p)\,U_{\mu}\,U_{\nu} - p\,g_{\mu\nu} +
\partial_{\mu}\,\varphi\,\partial_{\nu}\,\varphi -
\frac{1}{2}\,(\partial^{\alpha}\,\varphi\,\partial_{\alpha}\,\varphi
- 2\,V(\varphi))\,g_{\mu\nu}
\end{equation}

\noindent where $\rho$ and $p$ are the energy density and pressure of the
fluid, respectively. In the present model we assume the equation of state for
the fluid, $p=1/3\,\rho$. This assumption is justified since the models are
supposed to describe a phase of the Universe emerging from the Planck
era\cite{hawking}. The Einstein's equations for the metric (2.1) and the
energy momentum tensor (2.2) are equivalent to the Hamiltonian's equations
generated by the Hamiltonian constraint

\begin{equation}
H = -\frac{p_{\varphi}^2}{2\,a^3} + \frac{p_a^2}{12\,a} + 3\,a -
a^3\,V(\varphi) - \frac{E_0}{a} = 0,  \label{Hamiltonian}
\end{equation}

\noindent where $p_a$ and $p_{\varphi}$ are the momenta canonically
conjugated to $a$ and $\varphi$, respectively, $E_0$ is a constant arising
from the first integral of Bianchi identities, and obviously proportional to
the total matter energy of the models. The complete dynamics is governed by
Hamilton's equations

\begin{equation}
\left\{
\begin{array}{ll}
\dot{a} = \frac{p_a}{6\,a} \\
\dot{\varphi} = - \frac{p_{\varphi}}{a^3} \\
\dot{p_a} = -\frac{p_{\varphi}^2}{a^4} - 6 + 4\,a^2\,V(\varphi)\\
\dot{p_{\varphi}} = a^3\,V^{\prime}(\varphi)
\end{array}
\right. \           \label{syst1}
\end{equation}

\noindent where $\prime$ denotes derivative with respect to $\varphi$. We
have used the Hamiltonian constraint (\ref{Hamiltonian}) to simplify the
third of Eqs. (\ref{syst1}). The dynamical system (\ref{syst1}) has critical
points $E_{\varphi_0}$ in the finite region of the phase space whose
coordinates are

\begin{equation}
E_{\varphi_0}:    \varphi = \varphi_0,   a = a_0 = \sqrt{\frac{3}{2\,V(\varphi_0)}},
  p_a = 0,  p_\varphi = 0,         
\end{equation}

\noindent where $\varphi_0$ is solution of the equation $V^{\prime}(\varphi_0) = 0$.
The number of critical points
$E_{\varphi_0}$ is therefore equal to the number of extrema $\varphi_0$ of the
potential $V(\varphi)$. The energy associated with each critical point is given by

\begin{equation}
E_0 = E_{cr} = \frac{3\,a_0^2}{2}. 
\label{energy}
\end{equation}

\noindent Also, the existence of the critical points demands that
$V(\varphi_0)>0$; these critical points correspond to the configuration of
the Einstein Universe, with the respective cosmological constant given by
each $V(\varphi_0)$. The stability analysis is obtained by linearizing the
system (\ref{syst1}) about the critical points $E_{\varphi_0}$, resulting in
the constant matrix determining the linear system about each $E_{\varphi_0}$,
with the four eigenvalues

\begin{equation}
\lambda_{1,2} = \pm\,\sqrt{\frac{4\,V(\varphi_0)}{3}}, \, \,  \lambda_{3,4} =
\pm\,\sqrt{-V^{\prime\prime}(\varphi_0)}.
\end{equation}

\noindent From above we can see that if the particular extremum of
$V(\varphi)$ is a minimum, the corresponding critical point is a
saddle-center\cite{wiggins}. On the contrary, if
$V^{\prime\prime}(\varphi_0)<0$, the critical point is a pure saddle, with
four real eigenvalues.

The dynamical system (\ref{syst1}) admits invariant planes ${\cal{M}}_{\varphi_0}$ defined generically by

\begin{equation}
{\cal{M}}_{\varphi_0}: \varphi = \varphi_0, \, \,p_{\varphi} = 0.
\end{equation}

\noindent On ${\cal{M}}_{\varphi_0}$ the dynamics is governed by the
two-dimensional system

\begin{equation}
\left\{
\begin{array}{ll}
\dot{a} =  \frac{p_a}{6\,a} \\
\dot{p_a} =  - 6 + 4\,a^2\,V(\varphi_0).
\end{array}
\right. \ \label{syst2}
\end{equation}

\noindent The system (\ref{syst2}) is integrable, with Hamiltonian constraint
$H=\frac{p_a^2}{12\,a} + 3\,a-a^{3}\,V(\varphi_0) - \frac{E_0}{a} = 0$.
Introducing the conformal time $\eta$ by $d\,\eta=\frac{d\,t}{a}$, the
dynamical system (\ref{syst2}) and its associated Hamiltonian constraint
become regular at $a=0$, so that the dynamics can be continuously extended to
the region $a<0$\footnote{We note however that $a = 0$ corresponds to the
singularity of the curvature tensor and we will obviously restrict our
analysis to the physical region $a > 0$.} as showed in Fig. 1. The integral
curves represent closed FRW models, with radiation plus an effective
cosmological constant given by the value $V(\varphi_0)$. We remark that each
critical point $E_{\varphi_0}$ belongs to the respective invariant plane
${\cal{M}}_{\varphi_0}$. A straightforward analysis of the infinity of the
phase space shows the presence of pairs of critical points, corresponding to
the de Sitter solution, one acting as an attractor (stable de Sitter
configuration) and the other as a repeller (unstable de Sitter
configuration). Each pair of de Sitter attractor is associated to an
invariant plane ${\cal{M}}_{\varphi_0}$, with its corresponding effective
cosmological constant $V(\varphi_0)$. The scale factor $a(t)$ approaches the
stable de Sitter attractors as $a(t) \sim e^{\sqrt{V(\varphi_0)/3}\,t}$. The
stable de Sitter attractors define exits to inflation, and one of the
questions to be examined in this paper is the characterization of sets of
initial conditions for which one of them is attained. The existence of
critical points of saddle-saddle or saddle-center character, and of distinct
exits to inflation associated to the de Sitter attractors, is a striking
novelty in the dynamics of our models. Moreover, we will show that - due to
topology of the phase space of the models - the dynamics of the allowable
exits to inflation is highly complex. For example, sets of initial conditions
exist for which the exit to inflation is chaotic: arbitrarily small
fluctuations of initial conditions in these sets may change the final state
of the universe, not only from collapse to escape in the neighborhood of the
same invariant plane, but also from collapse/escape about the neighborhood of
one invariant manifold into collapse/escape about the neighborhood of another
invariant plane.

This complex dynamics and some of its physical applications will be discussed
in the next Sections, together with various numerical experiments showing the
above mentioned effects, for the case of two standard scalar field
potentials: the first with one extremum (minimum) only, and the second having
three extrema (two minima and one maximum).

\section{The Topology of Phase Space about the Critical Points}

Our starting point here is to linearize the Hamiltonian (\ref{Hamiltonian})
about its critical points.  As we have seen already these critical points are
of two types, a saddle-center if $V^{\prime\prime}(\varphi_0) > 0$ or a pure
saddle if $V^{\prime\prime}(\varphi_0) < 0$.  About the critical point, whose
coordinates are given by Eq. (2.5), the Hamiltonian may be expressed

\begin{equation}
H = - \frac{p_{\varphi}^2}{2\,a_0^3} - \frac{a_0^3}{2}\,V^{\prime\prime}(\varphi_0)\,(\varphi-\varphi_0)^2+\frac{p_a^2}{12\,a_0} - \frac{6}{a_0}\,(a-a_0)^2+\frac{1}{a_0}\,(E_{cr}-E_0) + {\cal{O}}(3) = 0, 
\end{equation}

\noindent where ${\cal{O}}(3)$ denotes higher order terms in the
expansion. In a small neighborhood of the critical point, these higher order
terms can be neglected and the motion is separable, with the partial energies

\begin{equation}
E_1 = \frac{p_{\varphi}^2}{2\,a_0^2} + \frac{a_0^4}{2}\,V^{\prime\prime}(\varphi_0)\,(\varphi-\varphi_0)^2,\;\;    E_2 = \frac{1}{12}\,p_a^2 - 6\,(a-a_0)^2             \label{energies}
\end{equation}

\noindent approximately conserved.  We have

\begin{equation}
- E_1 + E_2 + E_{cr}-E_0 \sim 0,       \label{la}
\end{equation}

\noindent where the quantity $E_{cr}-E_0$ is small. We consider first the
case of saddle-centers. We note that $E_1$ is always positive, due to
$V^{\prime\prime}(\varphi_0) > 0$, and is associated to rotational motion,
while $E_2$ has not a fixed sign and corresponds to hyperbolic motion
always. This is in accordance with a theorem by Moser\cite{Moser} stating
that there always exits a set of canonical variables such that, in a small
neighborhood of a saddle center, the Hamiltonian is separable into rotational
motion and hyperbolic motion pieces.  In this approximation, we note from
(\ref{energies}) that the scale factor $a(t)$ has pure hyperbolic motion, and
is completely decoupled from the scalar field pure rotational motion. The
general oscillatory behavior of the orbits are connected to the existence of
a manifold of unstable periodic orbits, associated to the saddle-center.

To see this, let us briefly describe the topology of homoclinic cylinders in
the phase space about a saddle-center. A detailed description is done in OSS
(cf. also references therein). Let us consider the possible motions in a
small neighborhood $N$ of the saddle-centers. In the case $E_2=0$ and $p_a =
0 = a-a_0$, the motions are unstable periodic orbits $\tau_{E_0}$ in the
plane $(\varphi,p_\varphi)$. Such orbits depend continuously on the parameter
$E_0$. For $E_2=0$, there is still the possibility $p_a= \pm\,(a-a_0)$, which
defines the linear stable $V_s$ and unstable $V_u$ one-dimensional manifolds,
which are tangent, at the critical point, to the separatrices $S$ of the
invariant plane associated to the saddle-center (cf. Fig. 1). The
separatrices are actually the nonlinear extension of $V_s$ and $V_u$. The
direct product of the periodic orbit $\tau_{E_0}$ with $V_s$ and $V_u$
generates, in the linear neighborhood $N$ of the saddle-center, the structure
of pairs of stable $(\tau_{E_0} \times V_s)$ and unstable $(\tau_{E_0} \times
V_u)$ cylinders. Orbits on the cylinders coalesce into the periodic orbit
$\tau_{E_0}$ for times going to $+\infty$ and $-\infty$, respectively, the
energy of the orbits being the same as that of the periodic orbit. The
nonlinear extension of the plane of rotational motion, where the linear
unstable periodic orbits reside, is a two-dimensional manifold, the {\it
center manifold}\cite{ivano}, of unstable periodic orbits of the system,
parametrized with the energy $E_0$. The intersection of the center manifold
with the energy surface $E_0$ is a periodic orbit of energy $E_0$, from which
two pairs of cylinders emanate, as in the linear case regime previously
described. From (\ref{la}) we can see that the intersection of the center
manifold with the energy surface $E_0=E_{cr}$ is just the critical point; for
$E_{cr}-E_0 < 0$, the energy surface does not intersect the center
manifold. It follows that the structure of homoclinic cylinders is present
only in the energy surfaces for which $E_{cr}-E_0 > 0$. In the case $E_2 \neq
0$ and $E_{cr}-E_0 > 0$ the orbits are restricted to infinite cylindrical
surfaces which, in a linear neighborhood $N$ of the saddle-center, are the
product of periodic orbits of the central manifold with small hyperbolae in
the plane $(a,p_a)$, which are obviously solutions of (\ref{energies}). We
remark that a picture of the these hyperbolae is given by the curves of
Fig. 1 contained in the small neighborhood $N$.

Now due to the non-integrability of the system, the extension of the
cylinders away from the periodic orbit is distorted and twisted, with
eventual transversal crossings of the unstable cylinder with the stable
one. These intersections will produce chaotic sets in the phase
space\cite{oss}\cite{ozorio}\cite{vieira}, analogously to the case of
breaking and crossing of homoclinic/heteroclinic connections in Poincare's
homoclinic phenomena\cite{berry}. This provides an invariant characterization
of chaos in the general relativistic dynamics of the models.

From the general chaotic behavior of the system, we will single out the
following aspect which is of physical interest for inflation. A general orbit
which visits the neighborhood $N$ is characterized by $E_1 \neq 0$, $E_2 \neq
0$. In this region the orbit has an oscillatory approach to the cylinders,
the closer as $E_2 \rightarrow 0$. The {\it partition} of the energy $|E_{cr}
- E_0|$, into the energies $E_1$ and $E_2$ of motion about the critical
point, will determine the outcome of the oscillatory regime into collapse or
escape to inflation (de Sitter attractor). For instance, initially expanding
models with energy $E_0$ will go out the oscillatory regime into collapse or
escape if the {\it partition} of $E_{cr} - E_0$ in $N$ is such that $E_2 < 0$
or $E_2 > 0$, respectively. However, the non-integrability of the system
(\ref{syst1}), with the consequent twisting and crossing of homoclinic
cylinders, will cause that this {\it partition} of energy is chaotic in
general, and will characterize a chaotic exit to inflation towards the de
Sitter attractor of the invariant plane associated to the saddle-center. In
other words, given a general initial condition of energy $E_0$, we are no
longer able to foretell in what of regions $I$ or $II$ about the
saddle-center (cf. Fig.1) the orbit will stay when it approaches the
saddle-center. Small fluctuations in $E_0$ or in the initial conditions in
these sets will change the outcomes of the orbits from collapse and escape
and vice-versa, characterizing a chaotic exit to inflation. We remark this
chaotic behavior will also set up associated to cylinders emanating from
unstable periodic orbits of the center manifold which are not in a linear
neighborhood of the saddle-center. This will be illustrated thoroughly in the
numerical experiments in the following section.

But this is not the whole story of the chaotic exit to inflation in the
present model, as we will see if a critical point of the saddle-type is also
present in phase space. In the case of a pure saddle, both $E_1$ and $E_2$
have not a fixed sign (note that $V^{\prime\prime}(\varphi_0) < 0$) and
correspond obviously to hyperbolic motion only. Instead of homoclinic
cylinders in this neighborhood, we have two sets of linear stable $V_s$ and
unstable $V_u$ manifolds emanating from the pure saddle, associated to the
pair of real eigenvalues (2.7). It is straightforward to see that one of the
sets, associated to the eigenvalues $\lambda_{1,2}$, is the linearization
(around the saddle point) of the separatrices of the invariant plane
containing this critical point. The non-linear extension of the second set
(corresponding to the eigenvalues $\lambda_{3,4}$) constitutes a homoclinic
curve that visits the nearest saddle-center, approaching a periodic orbit in
this neighborhood (cf. Fig. 6). The above-mentioned crossing of homoclinic
cylinders will reach the neighborhood of the invariant plane associated to
the pure saddle, producing also chaotic sets in this neighborhood.

The chaotic behavior of the system, associated to the exits to inflation,
will be the main object of the next Section.

\section{Chaotic Exits to Inflation}

In order to proceed in the numerical examination of the dynamics, we will
restrict ourselves to the two choices of the scalar field potential
$V(\varphi)$,

\begin{equation}
V_1(\varphi) = \Lambda + \frac{1}{2}\,m^2\,\varphi^2
\end{equation}

\noindent and

\begin{equation}
V_2(\varphi) = \Lambda + \frac{\lambda}{4}\,(\varphi^2 - \sigma^2)^2
\end{equation}

\noindent where $\Lambda$ is included as the vacuum energy of the scalar
field (inflaton field), $m$ is the mass of its expectation value, and
$\lambda$ and $\sigma$ are positive constants. The constant $\Lambda$ plays
the role of a cosmological constant. The usual matter content will be
represented by radiation. In the literature of inflation, an extensive use
has been made of both potentials\cite{inflation}. In the numerical
experiments performed here all calculations were made using the package {\it
Poincar\`{e}}\cite{ed}, where we enforce that the error of the Hamiltonian
never exceeds a given threshold of $10^{-10}$.

\subsection{Case $V_1(\varphi) = \Lambda + \frac{1}{2}\,m^2\,\varphi^2$}

The potential has only one extremum (a minimum) at $\varphi_0 = 0$, and the
phase space has only one critical point $P$ of saddle-center type, with
coordinates

\begin{equation}
a_0 = \sqrt{\frac{3}{2\,\Lambda}};\;
\varphi = \varphi_0 = 0;\; p_a = p_{\varphi} = 0,
\end{equation}

\noindent and energy $E_{cr} = \frac{9}{4\,\Lambda}$. The eigenvalues (2.7)
are given, respectively by $\pm\,\sqrt{\frac{4\,\Lambda}{3}}$ and
$\pm\,i\,m$, and they are associated to the hyperbolic motion in the plane
$(a,p_a)$ and rotational motion in the plane $(\varphi,p_\varphi)$, in a
neighborhood of $P$. Note that the eigenvalues corresponding to the
rotational motion depend on the mass of the scalar field. There is also one
invariant manifold ${\cal{M}}$ defined by $\varphi=p_{\varphi}=0$.

The phase space under consideration is not compact, and we will actually
identify a chaotic behavior associated to the possible asymptotic outcomes of
the orbits in this phase space, namely, escape to a de Sitter attractor at
infinity representing the inflationary regime, or collapse after a burst of
initial expansion.

Following the procedure of OSS, our objective here is first to analyze
numerically the behavior of the orbits, the initial conditions of which are
taken in a small neighborhood of a point on the separatrix in the invariant
manifold ${\cal{M}}$ ($\varphi=p_{\varphi}=0$). We assume that $\Lambda =
1.5$ and $m = 4.0$ such that $a_0=1.0$ and $E_{cr}=1.5$. We select a point
$S_0$ belonging to the separatrix ($E_0 = E_{cr} = 1.5$) with coordinates $a
= 0.4$, $p_a = 3.563818177$; around $S_0$ we construct a four dimensional
sphere in phase space with arbitrary small radius $R=10^{-2}, 10^{-3},
10^{-4}$, as the measure of the uncertainty in the initial conditions. The
initial conditions $(a,p_a,\varphi,p_\varphi)$ are then taken in energy
surfaces which have a non-empty intersection with this sphere, as evaluated
from the Hamiltonian constraint. It is easy to see that such energy surfaces
are those for which the range of the energy $E_0$ is in the interval
${\cal{D}} = (1.5 - \Delta\,E_0, 1.5 - \Delta\,E_0)$ about the critical
energy $E_0 = E_{cr} = 1.5$, with $\Delta\,E_0$ of the order of, or smaller
than $R$. Actually, these initial conditions represent initially expanding
models just after the singularity exhibiting small perturbations in the
scalar field. The numerical experiments revealed, as expected, two possible
outcomes: collapse or escape to the inflationary regime, depending on $E_0$
that varies from $1.5 - \Delta\,E_0$ to $1.5 + \Delta\,E_0$, with
$\Delta\,E_0$ of order $R$. Since the energy $E_0$ was chosen to be very
close to the energy of the separatrix, it is not difficult to prove\cite{oss}
that all orbits visit a small (of order of $R$) neighborhood $N$ of the
critical point before collapsing or escaping into inflation. The final state
depends crucially on the {\it partition} of the energy $|E_0-E_{cr}|$ into
the rotational motion mode and the hyperbolic motion mode, in the small
neighborhood $N$, so that, if $E_2 > 0$ the orbits escape, whereas collapse
is characterized by $E_2 < 0$. In Fig. 2 we illustrate the collapse and the
escape of 400 orbits initially in a sphere of radius $R = 10^{-4}$.
Nevertheless, for each radius $R$, there exists a non-null interval of energy
$\delta\,E^* = |E_{max} - E_{min}| \subset {\cal{D}}$ for which orbits have
an indeterminate outcome, that is, fluctuations in initial conditions of the
order of or smaller than $R = 10^{-3}, 10^{-4},...$ will the long time
behavior of an orbit from collapse to escape into inflation, and
vice-versa. Here $E_{min}$ and $E_{max}$ denote the values of the energy
$E_0$ above/below which all orbits escape/collapse, respectively. In this
sense we say that the exit to inflation is chaotic. In Fig. 3 this behavior
is showed for 300 orbits corresponding to a sphere of initial conditions with
radius $R = 10^{-4}$. This result of fundamental importance is an evidence of
the chaotic {\it partition} of the energy $E_0-E_{cr}$ into the energies
modes $E_1$ and $E_2$ when the orbits are in the small region around $P$. It
is worth to mention that the indeterminate outcome due to $\delta\,E^* \neq
0$ occurs only for $E_0-E_{cr} < 0$, as expected. This is the energy
condition for the presence of homoclinic cylinders. In the case $E_0-E_{cr}
\geq 0$ yields $E_2 > 0$ and all orbits escape. By determining numerically
the values of $E_{min}$ and $E_{max}$ for several values of $R$, we obtain
the scaling law, $\delta\,E^* = k\,R^2$, where a is constant depending on
$m$. We will discuss this dependence on $m$ elsewhere, but for $m=4$, we have
$k \approx 2.576$. An analogous scaling relation was obtained in OSS, in the
realm of anisotropic universes. We recall that the above relation is a
manifestation of chaos resulting from the crossing of unstable and stable
cylinders emanating from the periodic orbits of the center manifold.

The above chaotic behavior is not restricted to sets of initial conditions
infinitesimally close to the invariant manifold and whose orbits visit a
small neighborhood of $P$. In Fig. 4 we show the chaotic exit to inflation
occurring in a non-linear neighborhood of the saddle-center in which $a(t)$
and $p_a(t)$ oscillate, implying in the breakdown of the linear
approximation, since, according with Section III, the infinitesimal
neighborhood of $P$ displays only hyperbolic motions in the plane
$(a,p_a)$. In this non-linear regime about $P$ the scalar field degree of
freedom pumps rotational energy into the degree of freedom associated to the
gravitational scale factor. The initial conditions were obtained by
constructing a small sphere about a point outside the invariant manifold. All
orbits oscillate very close to a typical unstable periodic orbit of the
center manifold before collapsing or escaping to inflation. The set of
unstable periodic orbits are very special solutions that can not be exactly
attained by a physically relevant solution. This numerical experiment also
indicates the generality of the chaotic exit to inflation due to the {\it
partition} of the total energy into the rotational energy mode of the
periodic orbit and the hyperbolic mode. The latter fixes along which cylinder
(of collapse/escape) the motion will flow.

\subsection{Case $V_2(\varphi) = \Lambda + \frac{\lambda}{4}\,(\varphi^2 -
\sigma^2)^2$}

For the scalar field potential (4.2), three critical points are present on the
phase space. Two of them are saddle-centers whereas the third is a pure saddle
denoted, respectively, by $P_{\pm}$ and $P_0$ whose coordinates are

\begin{equation}
P_{\pm}:\,\, a_{\pm} = \sqrt{\frac{3}{2\,\Lambda}}; \varphi = \varphi_0 =
\pm\,\sigma; p_a = p_{\varphi} = 0,
\end{equation}

\begin{equation}
P_0:\,\, a_0 = \sqrt{\frac{3}{2\,\Lambda_{ef}}};
\varphi = \varphi_0 = 0; p_a = p_{\varphi} = 0,
\end{equation}

\noindent where $\Lambda_{ef} = \Lambda + \frac{\lambda}{4}\,\sigma^4$. The
energies of $P_{\pm}$ and $P_0$ are given by $E_{cr} = \frac{9}{4\,\Lambda},
\frac{9}{4\,\Lambda_{ef}}$, respectively. Since the Hamiltonian
(\ref{Hamiltonian}) is invariant under the change $\varphi \rightarrow
-\varphi$ and $p_{\varphi} \rightarrow -p_{\varphi}$, both critical points
$P_{\pm}$ are physically identical. Due to the fact that $\Lambda_{ef} >
\Lambda$, it follows $a_{\pm} > a_0$. The critical points are contained,
respectively, in the three invariant manifolds ${\cal{M}}_{\pm}$ and
${\cal{M}}_{0}$, defined by ($\varphi = \pm\,\sigma$, $p_{\varphi}=0$), and
($\varphi = 0$, $p_{\varphi}=0$). The phase portrait of the invariant
manifolds are schematically showed in Fig. 1. As we shall see in the
sequence, the coexistence of three critical points of distinct topological
nature produce very rich and complex dynamics resulting in several chaotic
exits to inflation. In the numerical experiments performed in the sequence,
we assume $\Lambda = 1.0$, $\lambda = 0.5$, $\sigma = \sqrt{3}$, which
produces $a_0 = 0.840168050$, $E_{cr} = 1.058823529$ and $a_{\pm} =
1.224744871$, $E_{cr} = 2.25$ that are the values of the scale factor of the
pure saddle and saddle-centers together with the corresponding energies,
respectively.

Analogous to the case $A$, we start the numerical study by examining the long
time behavior of orbits generated from initial conditions inside a sphere (as
before the radius $R$ is of order $10^{-2}$, $10^{-3}$,..., etc) about a point
on the separatrices of one of the invariant manifolds ${\cal{M}}_{\pm}$. These
orbits represent universes with radiation and effective cosmological constant
$V(\pm\,\sigma)=\Lambda$ with fluctuations of the scalar field about the
symmetry-breaking scale $|\varphi|=\sigma$. The dynamics is similar to the one
analyzed in the case $A$ for initial conditions sets taken in a neighborhood
of the invariant manifold of the saddle-center, with a chaotic exit to
inflation of the same type as showed in Fig. 3. This study displays only the
dynamics of case B occurring in a region of phase space close to the invariant
manifold of the saddle-center. The dynamical aspects due to the existence of
another critical points of the pure saddle nature have not been evidenciated,
as well as the dynamics in the region between the two invariant manifolds.
For instance, the extension of the center manifold and its associated
structure of homoclinic cylinders will permeate the neighborhood of the pure
saddle invariant plane ${\cal{M}}_{0}$, producing the complex dynamics to be
discussed next.

Now we proceed by taking initial conditions near the invariant manifold
${\cal{M}}_0$ associated to the pure saddle. The idea is, again, to select a
point on ${\cal{M}}_0$ and construct a small sphere of initial conditions
with radius $R$ that represents the uncertainty about the point under
consideration due to fluctuations around the local maximum of the
potential. Depending on the energy $E_0$ collapse and escape to inflation
take place. Nevertheless, it can be showed numerically that there always
exist an interval of energy $\delta\,E^* = |E_{max} - E_{min}|$ for each
radius $R$, assumed sufficiently small as $R=10^{-3}$, $10^{-4}$,
$10^{-5}$,..., etc, in which the boundaries of collapse and escape to the de
Sitter configuration are chaotically mixed.  Again $E_{min}$ and $E_{max}$
denote the values of the energy $E_0$ for above/below which all orbits
escape/collapse, respectively. The chaotic exits to inflation occur for the
energy inside the interval $\delta\,E^*$, and are a direct consequence of the
nonintegrability of the dynamics between the saddle-centers and the pure
saddle. We recall that the twisting and crossing of homoclinic cylinders
emanating from periodic orbits of the center manifold extend to the region of
phase space between the saddle-centers and the pure saddle reaching the
neighborhood of ${\cal{M}}_0$. The interplay of cylinders in a neighborhood
of the pure saddle, and the consequent several chaotic exits to inflation,
can be revealed more clearly by the following experiments.  Consider now a
point lying on the separatrix $S$ of ${\cal{M}}_0$ and whose coordinates are
$a = 0.4$, $p_a = 2.756570759$, $\varphi = p_\varphi = 0$. Choosing the
radius $R$ sufficiently small, all orbits remains close to ${\cal{M}}_0$
until they reach a region of same order of $R$ around the pure saddle. From
this region, orbits will collapse or escape into the de Sitter attractor
associated to the invariant plane of pure saddle (with $\Lambda_{ef} =
\Lambda + \lambda\,\sigma^4/4$), depending on the energy $E_0$. However,
there also exists a domain $\delta\,E^*$ for which the outcome of orbits is
chaotic. According with Fig. 5, for a given energy inside the chaotic domain
$\delta\,E^*$, we note three types of orbits. Type $I$ orbits approach to the
pure saddle from which some collapse and some escape to inflation. Indeed, in
this linear region about the pure saddle, the {\it partition} of $|E_0 -
E_{cr}|$ into the hyperbolic modes energies $E_1$ and $E_2$ is completely
indeterminate so that we are not able to foretell which orbit will collapse
or escape once the initial conditions are generated. Type $IIa$ orbits visit
the neighborhood of the saddle-center, oscillate to follow with
collapse/escape, whereas type $IIb$ return to the neighborhood of the pure
saddle to proceed with collapse/ escape. In these situations the {\it
partition} of $|E_0 - E_{cr}|$ into the hyperbolic and rotational energies
modes around the saddle-center, and in the hyperbolic energy modes associated
with the pure saddle are chaotic. In Fig.  6, we refine the numerical
experiment in such a way to select only Type $IIb$ orbits. By projecting them
in the plane $\varphi,p_\varphi$, the approach to the homoclinic trajectory
appears. Therefore, in the same sense that orbits showed in Fig. 4 approached
a given unstable periodic orbit of the center manifold, the orbits of Fig. 6
approach the {\it homoclinic orbit}.

Another chaotic exit to inflation is obtained if we consider a point of
coordinates $a = 0.4$, $p_a = 4.427039907$, $\varphi = p_\varphi = 0$
($E_0 = 2.058823529$) on a trajectory (not the separatrix) lying
entirely on ${\cal{M}}_0$. For a given energy inside the chaotic domain the
initial conditions generated about this point evolve to one of the
saddle-centers, perform some oscillations in its neighborhood (the scale
factor $a$ and its canonical momentum $p_a$ also oscillate) to collapse or to
escape to inflation, as showed in Fig. 7. In this case, the orbits have
approached to a periodic orbit of the center manifold identically as showed in
Fig. 4. 

Finally, the chaotic behavior of the system described above, associated to
the several exits to inflation, can be summarized as follows:

\begin{enumerate} 

\item Small fluctuations of initial conditions taken on chaotic sets in a
neighborhood of the invariant plane associated to a saddle-center will change
one of the following asymptotic outcomes into another of the remaining ones:

\begin{itemize}

\item visit the neighborhood of the saddle-center and escape to inflation,
towards the de Sitter attractor associated to the invariant plane of the
saddle-center;

\item visit the neighborhood of the saddle-center, then visit a neighborhood
of the pure saddle, and escape to inflation, towards the de Sitter attractor
associated to the invariant plane of the pure saddle;

\item visit the neighborhood of the saddle-center and collapse;

\item visit the neighborhood of the saddle-center, then visit the neighborhood
of the pure saddle, and collapse.

\end{itemize}

\item Analogously, small fluctuations of initial conditions taken on chaotic
sets in a neighborhood of the invariant plane associated to a pure saddle will
change one of the following asymptotic outcomes into another of the remaining
ones:

\begin{itemize}

\item visit the neighborhood of the pure saddle and escape to inflation,
towards the de Sitter attractor associated to the invariant plane of the pure
saddle;

\item visit the neighborhood of the pure saddle, then visit a neighborhood of
the saddle-center, and escape to inflation, towards the de Sitter attractor
associated to the invariant plane of the saddle-center;

\item visit the neighborhood of the pure saddle and collapse;

\item visit the neighborhood of the pure saddle, then visit the neighborhood
of the saddle-center, and collapse.

\end{itemize}
\end{enumerate}

\noindent We recall that, in any of the exits to inflation, the scalar field
will be frozen in one of vacuum states (with its corresponding vacuum energy
playing the role of the cosmological constant for that exit) associated to one
of the extrema of the scalar field potential $V(\varphi)$.

\section{Final Remarks and Conclusions}

In this paper we have discussed the dynamics of closed
Friedmann-Robertson-Walker models which may provide a description of
preinflationary stages of the universe and its exit to inflation. The basic
physical ingredients of the models are radiation plus a scalar field
minimally coupled to the gravitational field. The energy momentum tensor of
the scalar field is split into a cosmological constant-type term
(corresponding to the vacuum energy of the scalar field), plus the energy
momentum tensor of the spatially homogeneous expectation value of the scalar
field. This simple configuration, with two effective degrees of freedom,
presents a complex dynamics. The basic features of the dynamics result from
the presence of saddle-center and pure saddle critical points in the phase
space of the system. In our model, the critical points are associated to
extrema of the scalar field potential, a minimum and a maximum corresponding,
respectively, a saddle-center and pure saddle. Each critical point is related
to an invariant plane of the dynamics and to a de Sitter attractor. The scale
factor approaches the de Sitter attractors exponentially, defining exits to
inflation, one for each critical point. The region of phase space about a
saddle-center has the structure of homoclinic cylinders, emanating from the
center manifold of unstable periodic orbits, resulting in that a general
orbit has an oscillatory behavior in the neighborhood of the
saddle-center. Due to the non-integrability of the system, the extension of
homoclinic cylinders away from the periodic orbit is distorted and twisted,
with eventual transversal crossings of the unstable cylinders with the stable
ones. These intersections produce chaotic sets in phase space, in a manner
analogous to the breaking and crossing of homoclinic/heteroclinic curves in
Poincar\'{e}'s homoclinic phenomena, and provide a topological
characterization of chaos in the general relativistic dynamics of the
model. As we have showed, this phenomena extends to the neighborhood of the
nearest pure saddle invariant plane, producing also chaotic sets of initial
conditions in the region of phase space laying between the invariant
planes. A physically relevant manifestation of chaos is the chaotic exit to
inflation through one of the de Sitter attractors present in phase
space. Small fluctuations of initial conditions taken on chaotic sets change
drastically the long time behavior of orbits, with all possibilities listed
at the end of Section IV. This is a fundamental result, illustrated
extensively in Figs. 3, 4, 5, 6 and 7. For instance, a typical chaotic exit
to inflation is realized when orbits, emanating from a small ball about a
point on the separatrix of one the invariant manifolds associated to the
saddle-centers reach a linear neighborhood of saddle-center to proceed with
collapse or escape to inflation. In this case, the chaotic domain is
characterized by the gap of energy $\delta\,E^*$ in which collapse and escape
are possible, implied by small fluctuations in initial
conditions. Nevertheless, the chaotic exit to inflation seems to be a general
feature of the system. As made evident in the text, there is always a gap of
energy for which orbits initially in a small sphere can visit a non-linear
region about the saddle-center to evolve afterwards to collapse or
escape. The new and important effect is the oscillation of the scale factor
induced by the scalar field, or, equivalently, due to the approach to a given
unstable orbit of the center manifold.  Considering a model with symmetry
breaking potential three critical points, two saddle-centers and one pure
saddle are present and the dynamics between them produces several chaotic
exits to inflation, which are related to the extension of the cylindrical
structure to this region. Therefore, we have showed numerically that there
always exist a gap of energy in which orbits initially close to the invariant
manifold of the pure saddle, for instance, can visit a small neighborhood of
one of the saddle-centers to $(i)$ collapse/escape, or $(ii)$ return to the
pure saddle to collapse/escape.

Finally, an interesting perspective of this work is the possibility of the
physical distinction between the exits to inflation, namely, whether the exit
occurred towards the saddle-center de Sitter attractor or towards the pure
saddle de Sitter attractor. This possibility is based on the growing of a
selected spectrum of Fourier components of inhomogeneous perturbations, due a
resonance mechanism generated by the oscillations of the scale factor, as
already pointed out previously\cite{oss}. Indeed, as we have seen, the scale
factor $a(t)$ and the scalar field $\varphi(t)$ oscillate, as the orbit
visits a neighborhood of the saddle-center, with frequency determined by the
unstable periodic orbit approached (see Fig. 4, for instance). We remark that
initial conditions always exist such that the oscillations take an arbitrary
fixed time before collapse or escape to a de Sitter phase. Therefore,
inhomogeneous scalar field perturbations and/or matter perturbations in this
gravitational background will have a selected spectrum of Fourier components
amplified by a mechanism of resonance with the oscillations, the
amplification occurring for the particular Fourier components having periods
approximately equal to an integer times the period of the periodic orbit
approached. Even if the universe inflates afterwards the relative rate of
amplitudes produced after the resonance amplification would be maintained as
an imprint in the ``initial espectrum" of density fluctuations. This
mechanism however is absent in the case of the exit towards the pure saddle
de Sitter attractor, since no oscillations appear when the orbit visits the
neighborhood of the pure saddle before escaping. The two cases can in
principle be observationally distinguished, based on restrictions imposed by
observations in the initial espectrum of density fluctuations. If the exit to
inflation occurred via a saddle-center de Sitter attractor the resonance
amplification mechanism referred to above will give rise to a
non-flat\cite{linde} ``initial espectrum" of density
fluctuations\cite{os}. The above analysis obviously excludes orbits of type
$IIb$.

\section{Acknowledgements}

The authors are grateful to CNPq and FAPERJ for financial support.

\newpage

\section*{Figure Captions}

Fig. 1 Phase portrait of the invariant manifold $\varphi = \varphi_0$,
$p_\varphi = 0$. The orbits represent homogeneous and
isotropic universes with radiation and cosmological constant.

\vspace{1cm}

Fig. 2 (a) Collapse of 100 orbits with energy $E_0 = 1.49999994$ initially in
a sphere of radius $R = 10^{-4}$ about the point $a = 0.4, p_a = 3.563818177$
on the separatrix of the invariant manifold associated to the
saddle-center. (b) Escape to inflation of 100 orbits with energy $E_0 =
1.499999999$ initially in a sphere of radius $R = 10^{-4}$ about the same
point.

\vspace{1cm}

Fig. 3 (a) Chaotic exit to inflation of 300 orbits with energy $E_0 =
1.499999983$ initially in a sphere of $R = 10^{-4}$ about the same point of
Fig. 2. (b) Three dimensional view of the region near the saddle-center. Note
the oscillations of the orbits in this region. (c) Projection of orbits near
the sphere of intitial conditions in the plane $(\varphi, p_{\varphi})$. The
strip in black indicates orbits that escape to de Sitter configuration, while
those in gray correspond to orbits that collapse. (d) A small strip of
Fig. 3(c) ($-0.00004 \leq \varphi \leq -0.00002$) is magnified, and repeats
the same pattern indicating a fractal structure.

\vspace{1cm}

Fig. 4 (a) Chaotic exit to inflation of 30 orbits with energy $E_0 =
1.3660351$ initially inside a ball of radius $ = 10^{-8}$ about the point
with coordinates $a = 0.560834374$, $p_a = 2.857528660$, $\varphi =
0.106525696$, $p_\varphi = 0.251541127$, close to the invariant manifold of
the saddle-center. (b) These orbits approach an unstable periodic orbit of
the center manifold in a small, but not infinitesimal, neighborhood of
saddle-center, in such a way that the scale factor $a(t)$ as well as $p_a(t)$
oscillate several times before collapsing or escaping. (c) Projection of the
same orbits in the plane $(\varphi,p_a)$ showing that the frequency of the
motion of $p_a$ is twice the frequency of the motion in $\varphi$.

\vspace{1cm}

Fig. 5 Chaotic exit to inflation (case of symmetry breaking potential) of 100
orbits with energy $E_0 = 1.058823529$ evolving from a sphere of $R =
10^{-7}$ about a point with $a = 0.4$, $p_a = 2.756570760$ on the separatrix
of the invariant manifold associated to the pure saddle point. The orbits
remain close to the invariant manifold until they arrive to the small
neighborhood of the pure saddle. Type $I$ orbits collapse or escape to
inflation after the approach to the pure saddle. Type $II$ orbits are
directed towards one of the saddle-centers and, after some oscillations,
either escape/collapse (type $IIa$), or return to a neighborhood of the pure
saddle (type $IIb$) to collapse/escape.

\vspace{1cm}

Fig. 6 (a) Chaotic exit to inflation of 100 orbits of type $IIb$ with $E_0 =
1.058825026$. Note the approach of these orbits to the homoclinic orbit
extending from the pure saddle to the saddle-center. (b) Zoom of the region
near the pure saddle showing the chaotic exit to inflation.

\vspace{1cm}

Fig. 7 Chaotic exit to inflation of 60 orbits with energy $E_0 =
1.623311538$, and initial conditions taken about the point, $a = 0.4$, $p_a =
4.427039909$, $\varphi = 0$, $p_\varphi = 0$. This point belongs to the
invariant manifold of the pure saddle point, but not on the separatrix. The
orbits visit the neighborhood of one of the saddle-centers and perform some
oscillations before the collapse/escape.

\newpage

\epsfysize=20cm
\centerline{\epsfbox{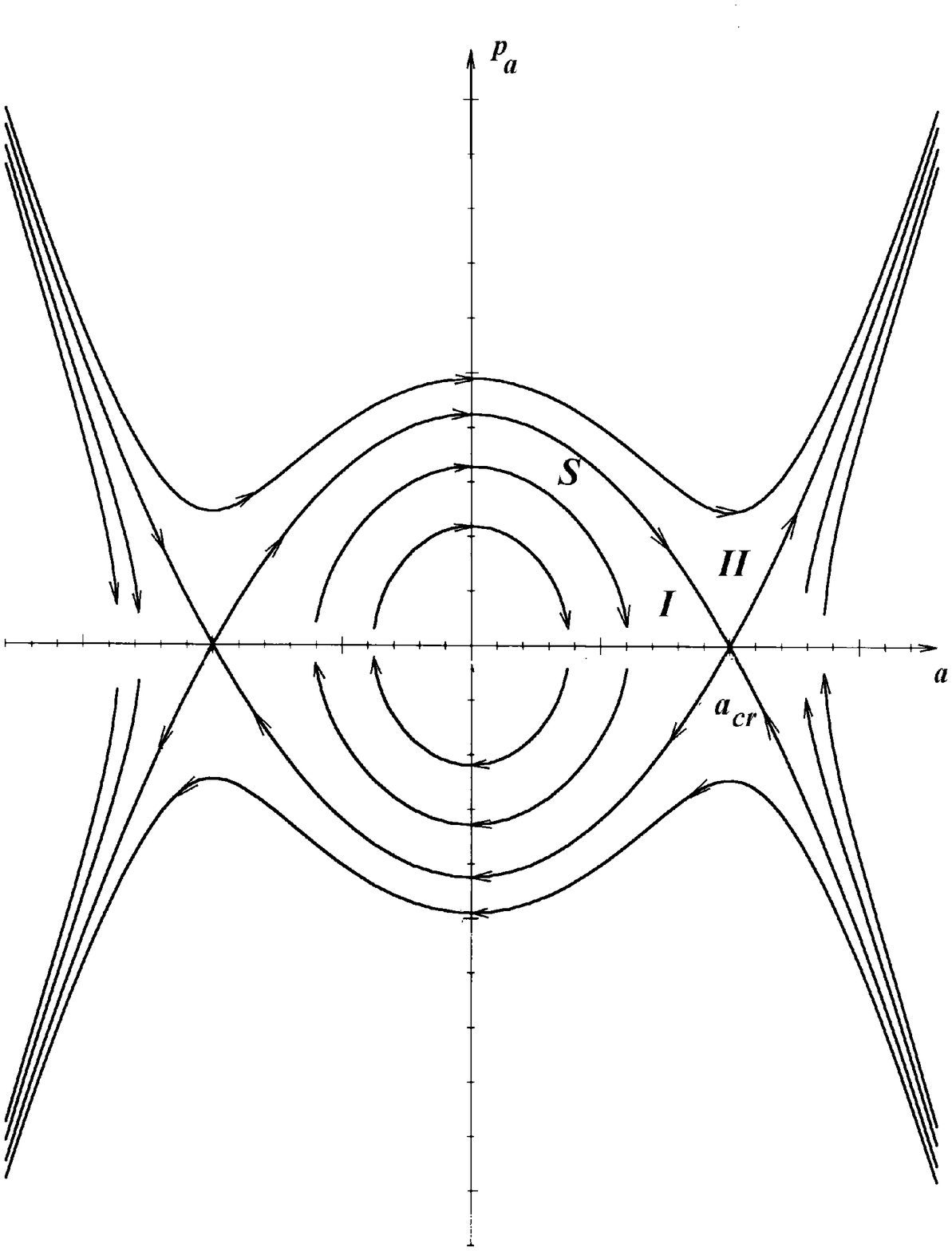}}

\newpage

\epsfysize=20cm
\centerline{\epsfbox{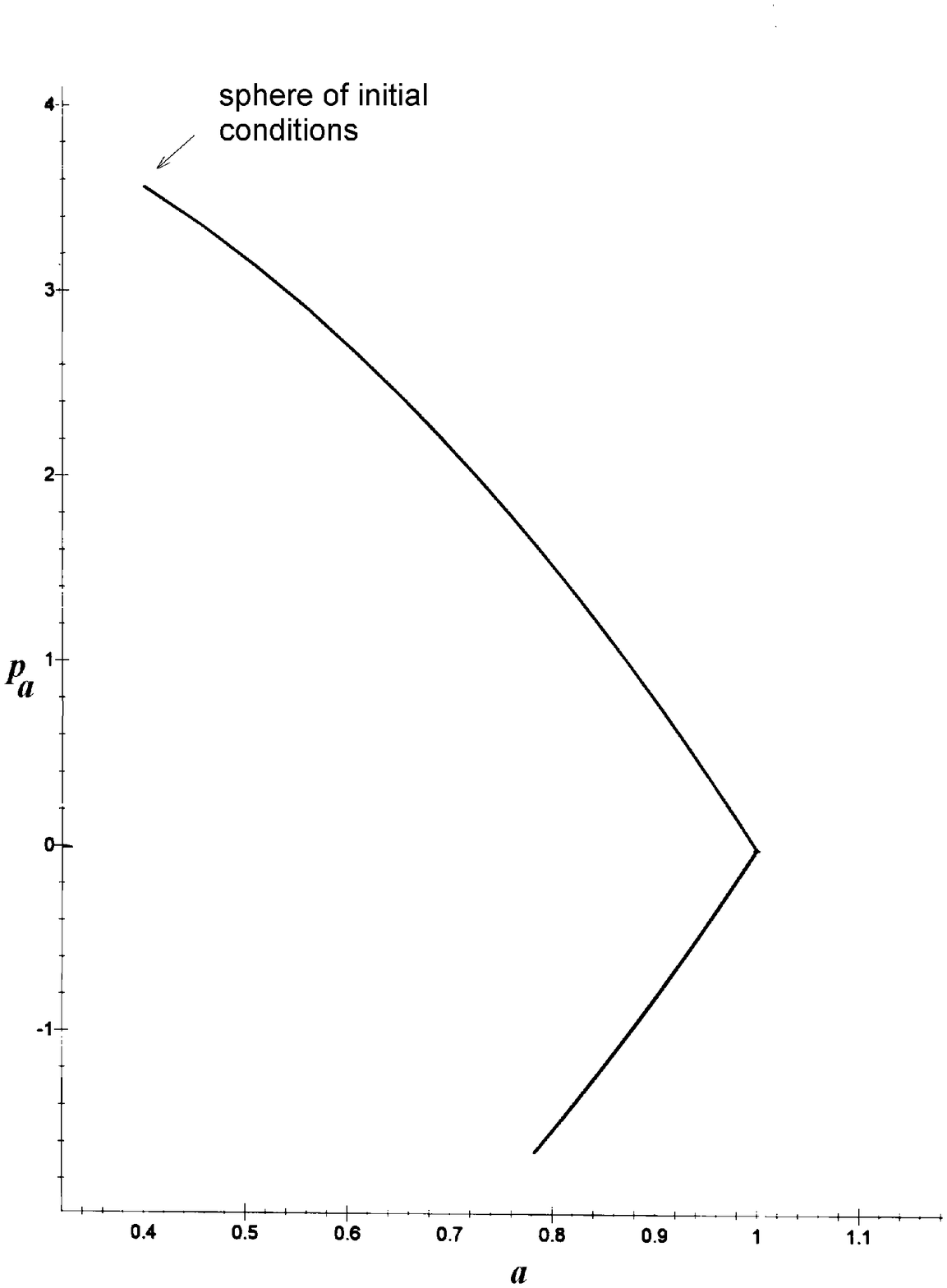}}

\newpage

\epsfysize=20cm
\centerline{\epsfbox{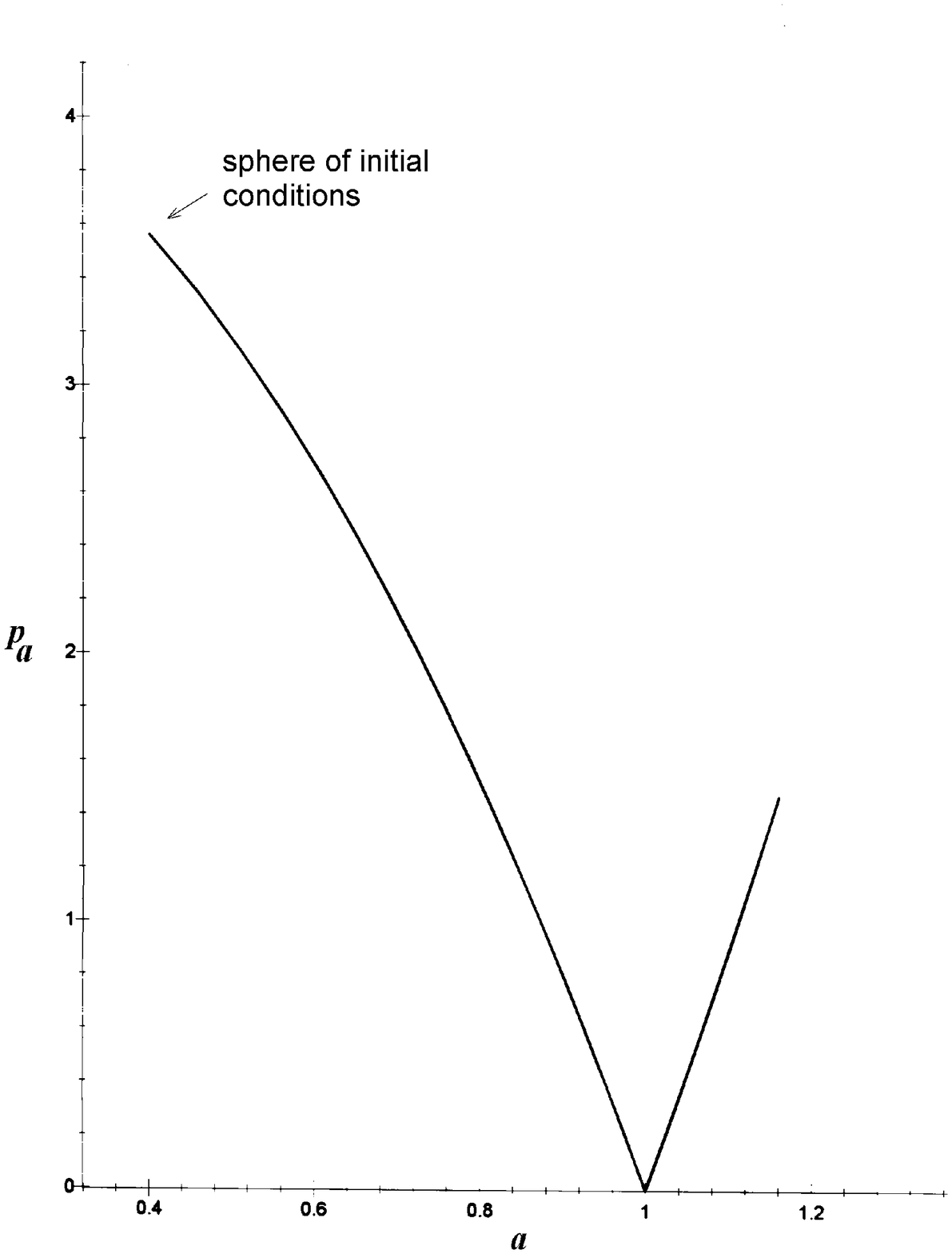}}

\newpage

\epsfysize=20cm
\centerline{\epsfbox{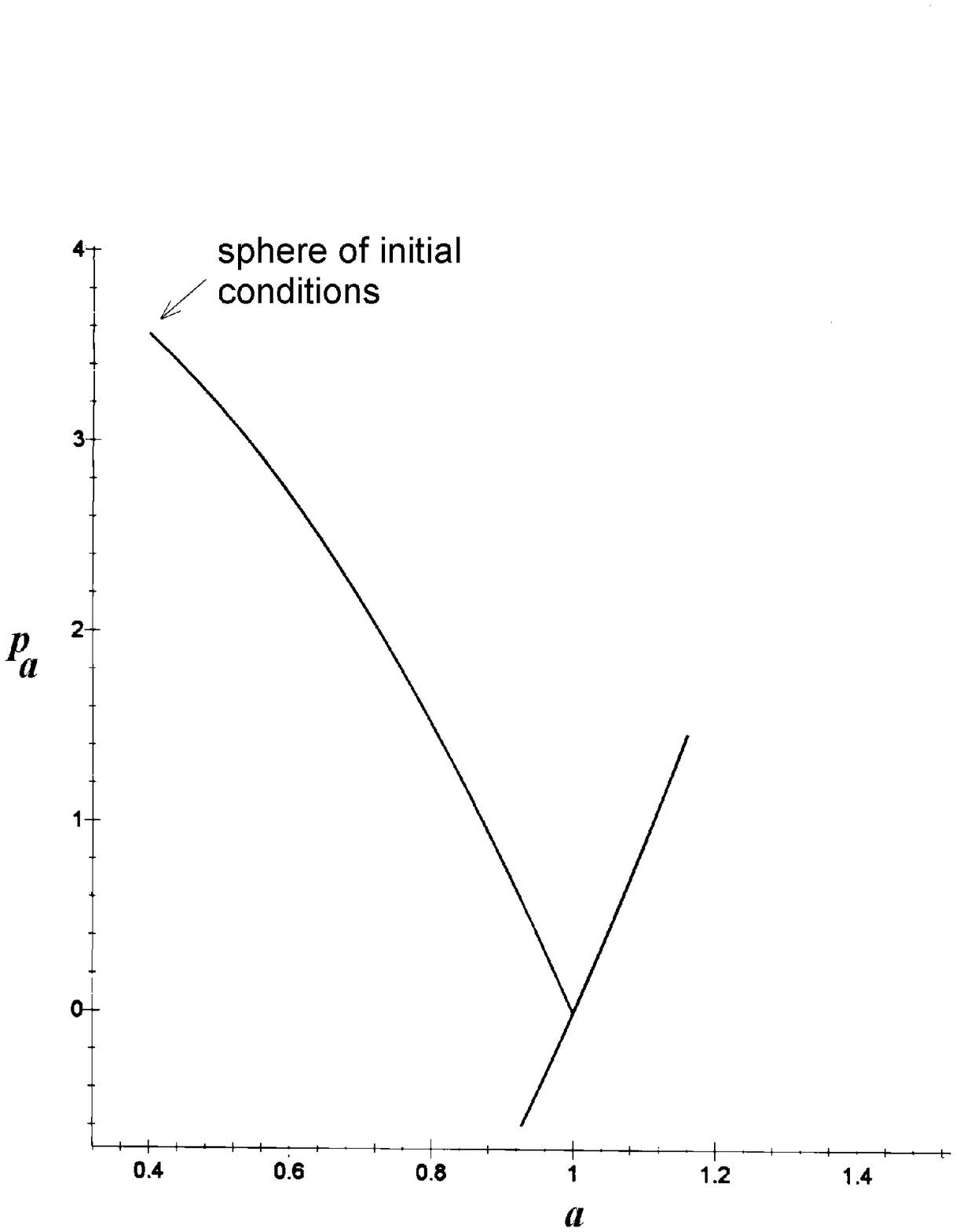}}

\newpage

\epsfysize=20cm
\centerline{\epsfbox{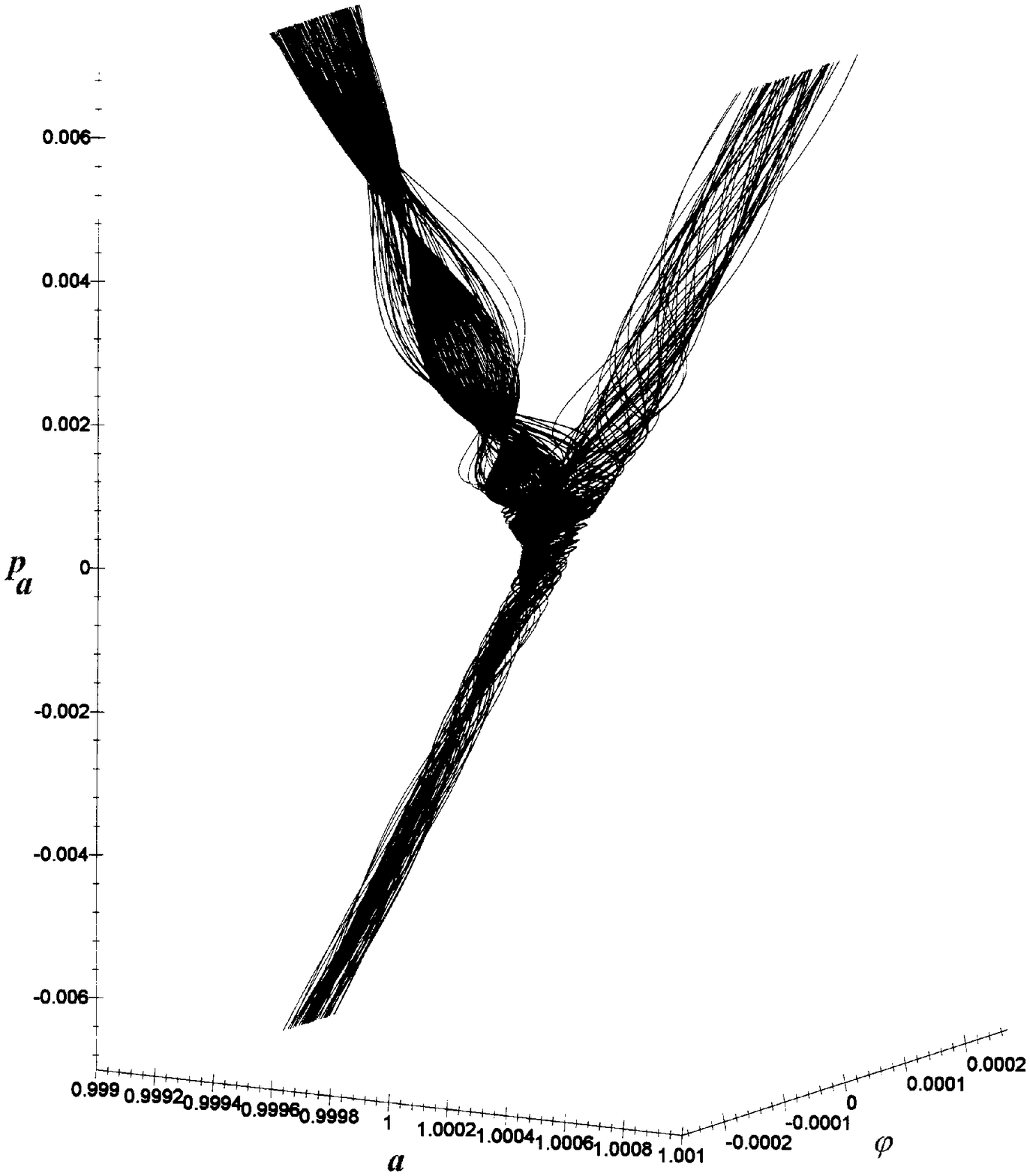}}

\newpage

\epsfysize=20cm
\centerline{\epsfbox{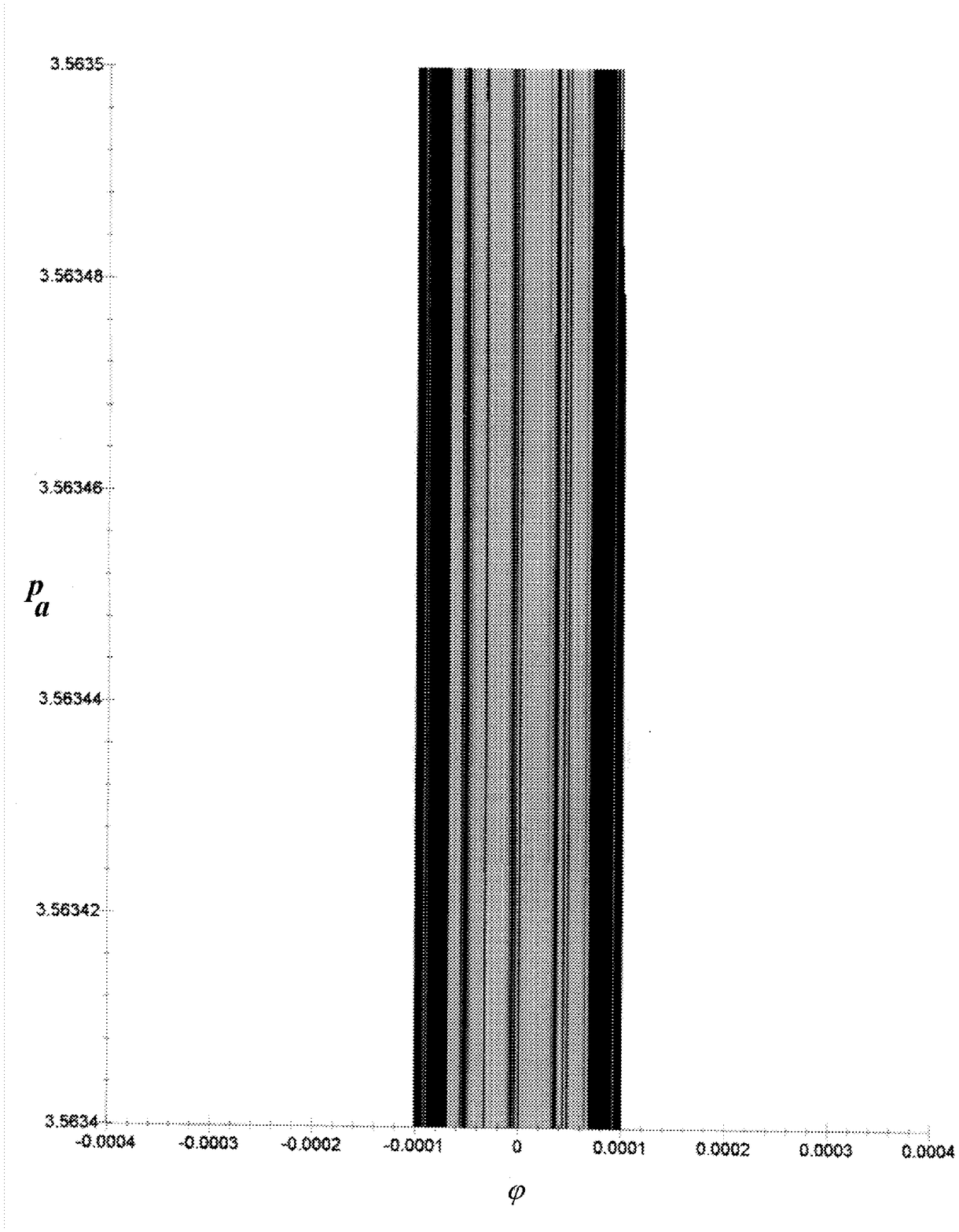}}

\newpage

\epsfysize=20cm
\centerline{\epsfbox{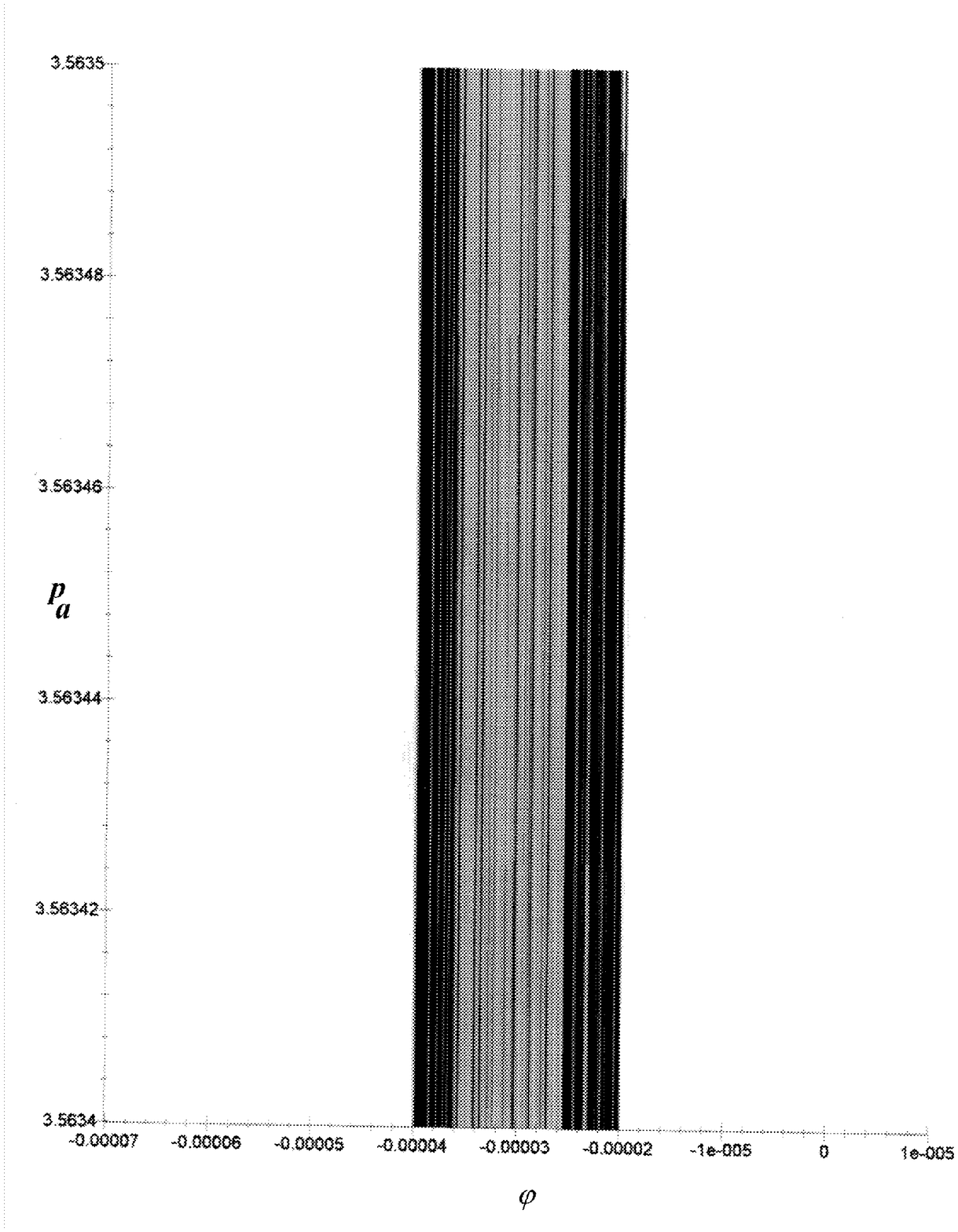}}

\newpage

\epsfysize=20cm
\centerline{\epsfbox{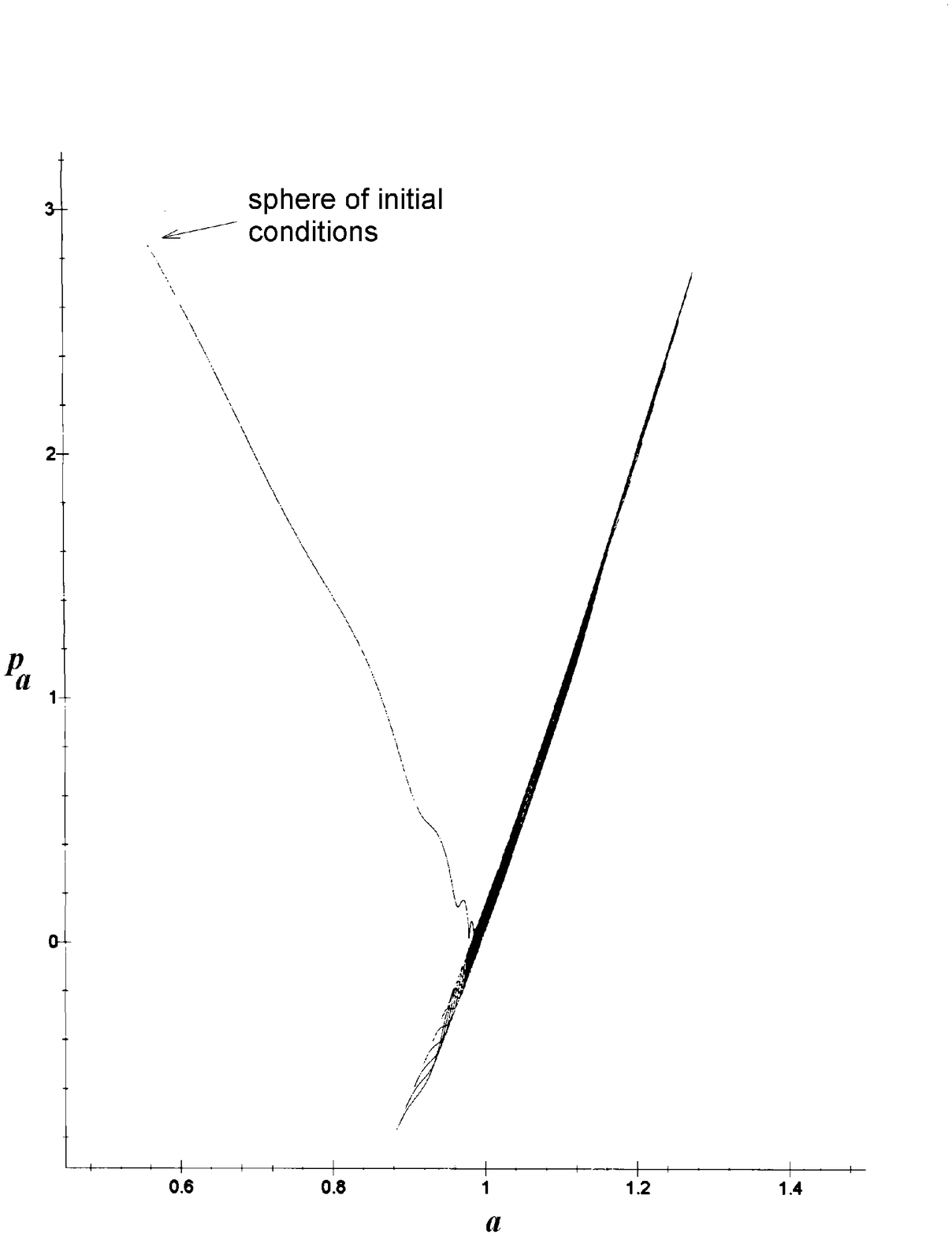}}

\newpage

\epsfysize=20cm
\centerline{\epsfbox{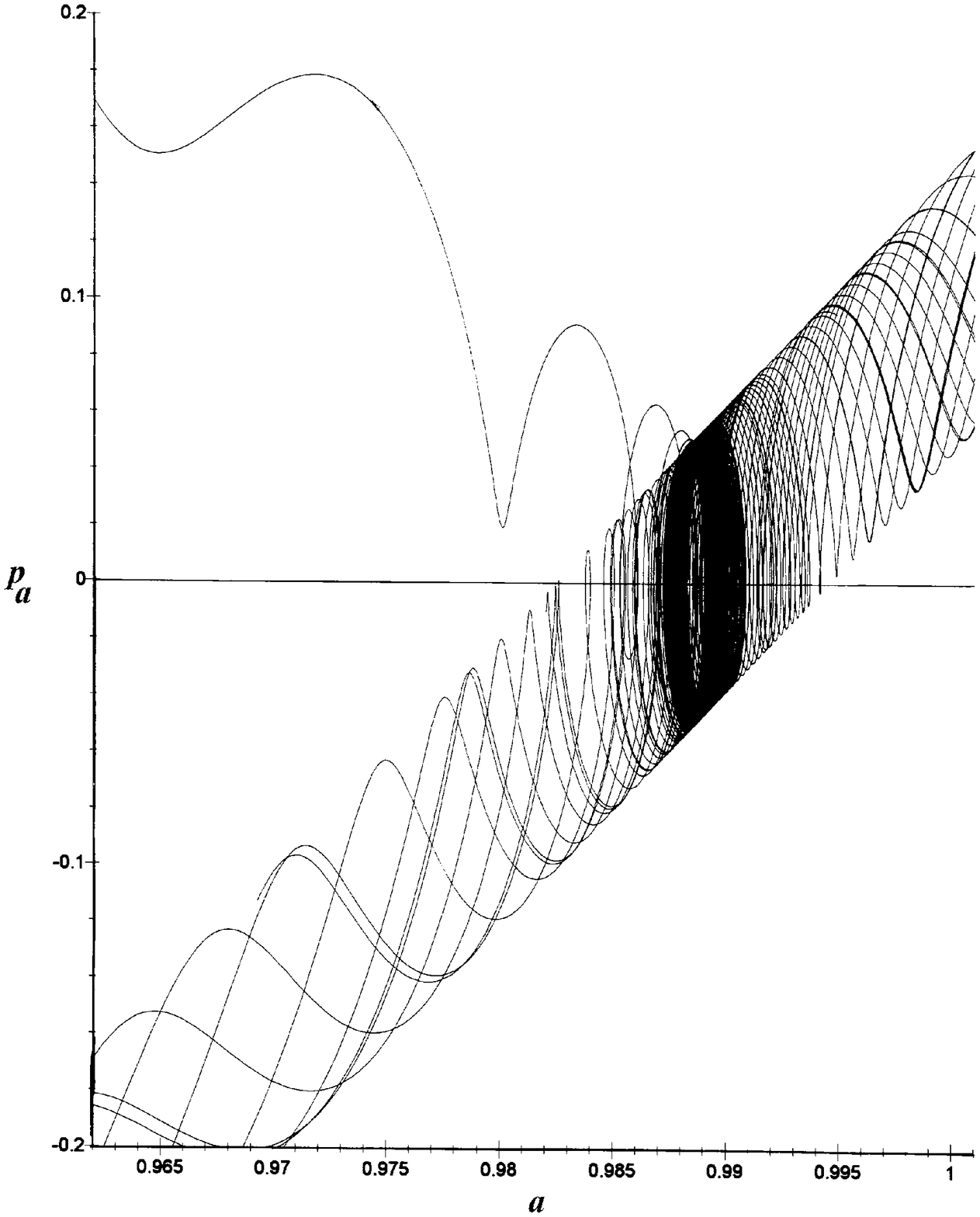}}

\newpage

\epsfysize=20cm
\centerline{\epsfbox{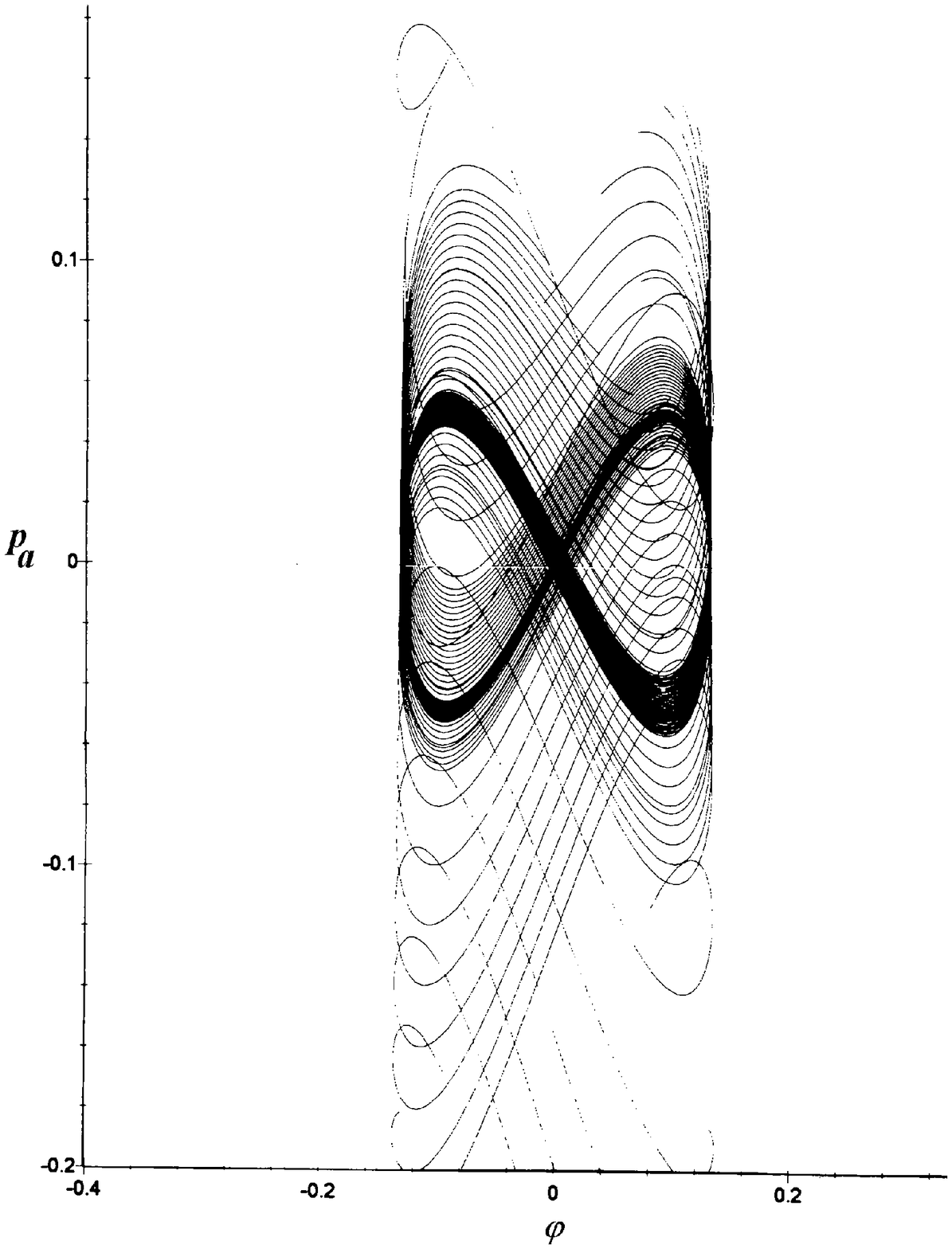}}

\newpage

\epsfysize=20cm
\centerline{\epsfbox{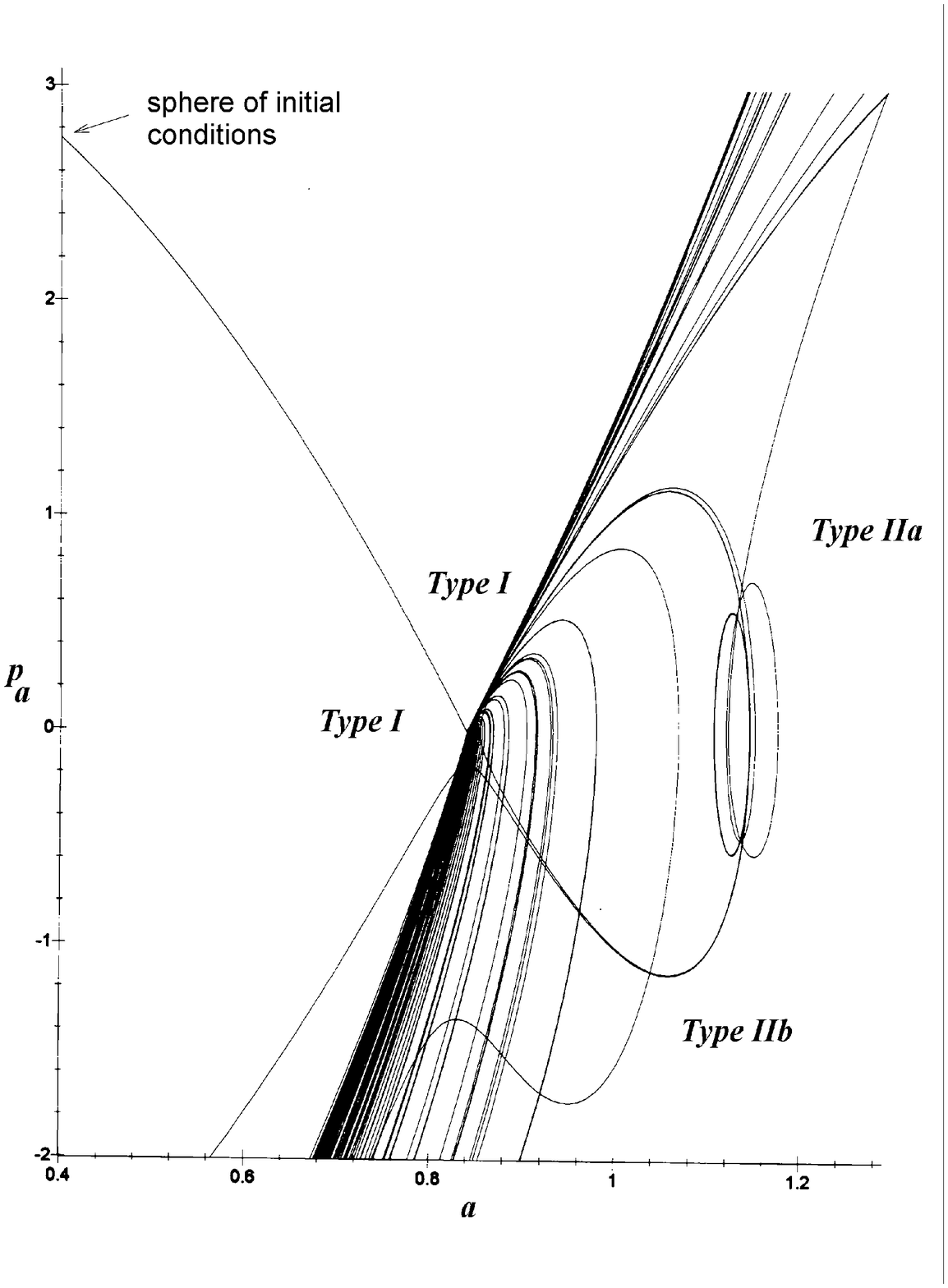}}

\newpage

\epsfysize=20cm
\centerline{\epsfbox{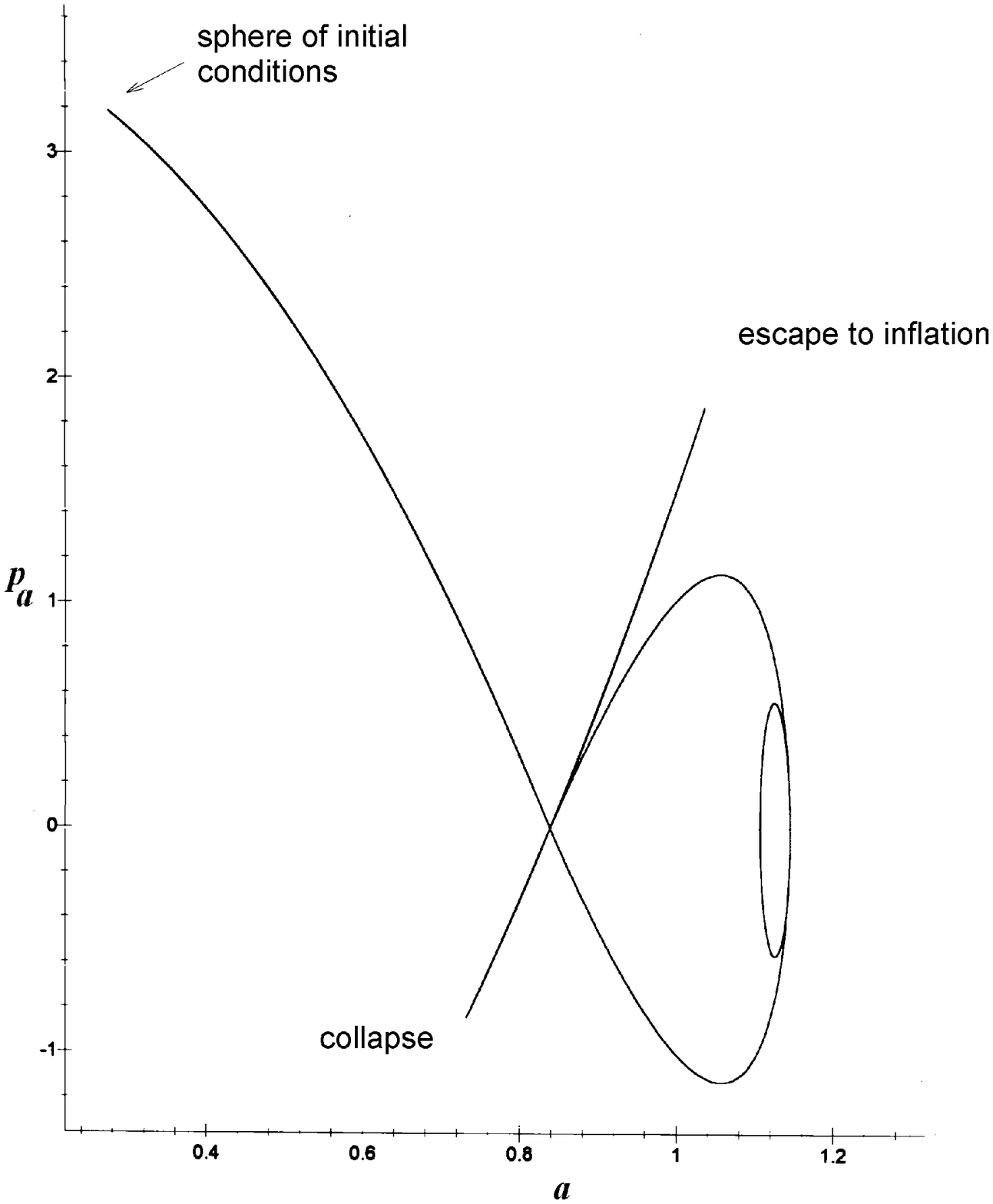}}

\newpage

\epsfysize=20cm
\centerline{\epsfbox{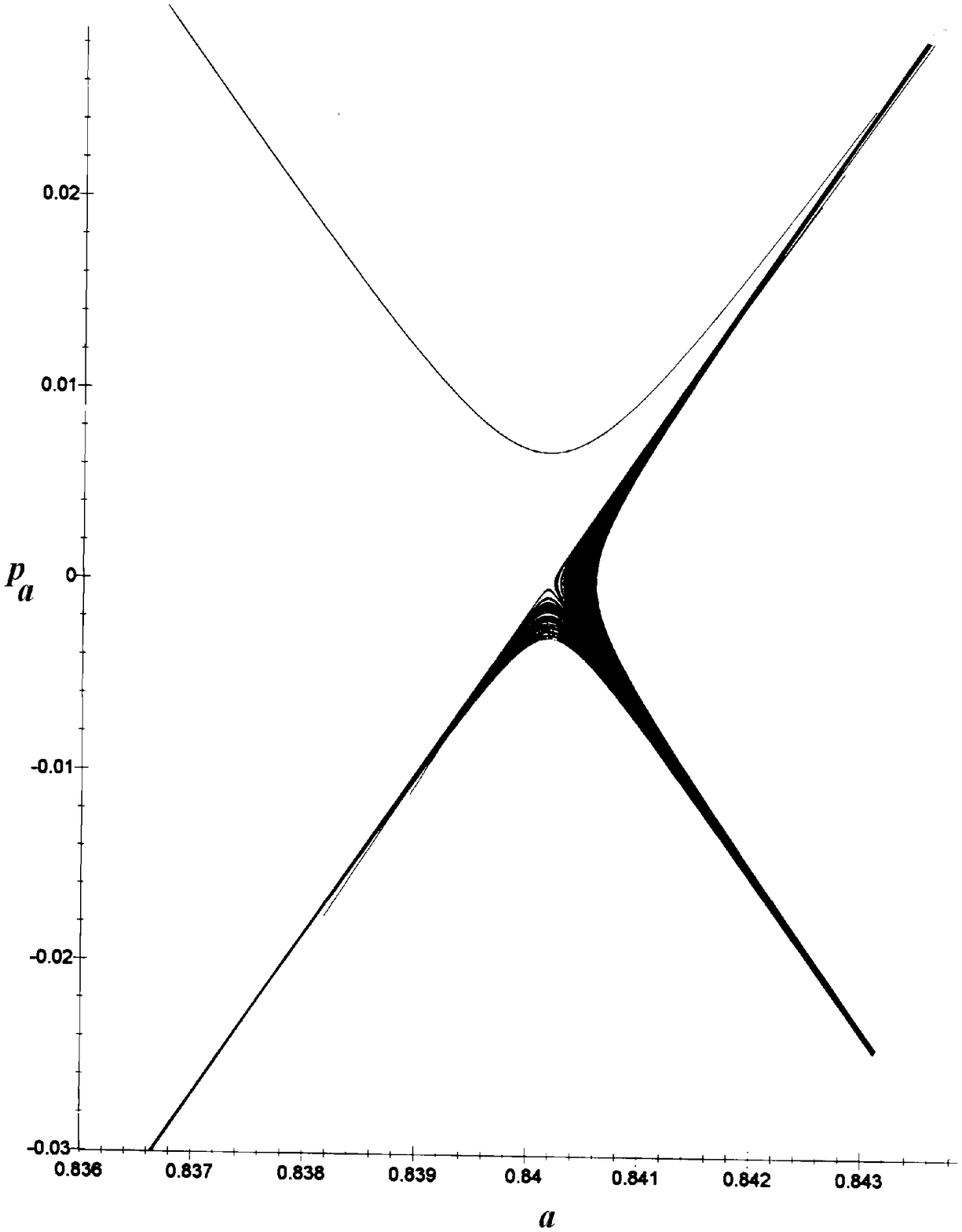}}

\newpage

\epsfysize=20cm
\centerline{\epsfbox{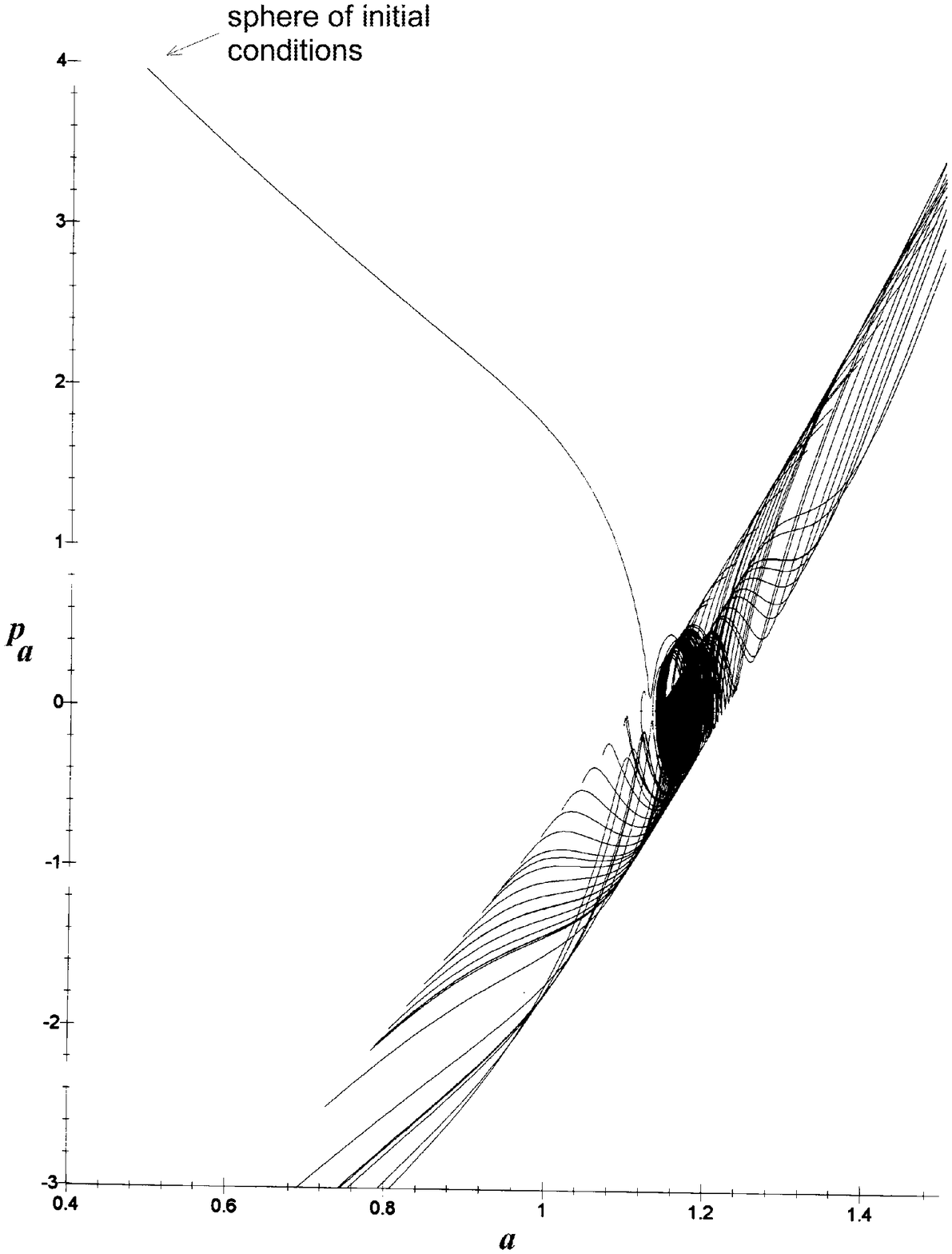}}


\begin{thebibliography}{99}

\bibitem{inflation} Edward W. Kolb and Michael S. Turner, {\it The Early
Universe}, Frontiers of Physics, Addison-Wesley Publishing Company (1990).

\bibitem{wald} R. M. Wald, Phys. Rev. D {\bf 28}, 2118 (1983).

\bibitem{starob} A. A. Starobinski, JETP lett., {\bf 37}, 66 (1983).

\bibitem{calza} E. Calzetta and E. El Hasi, Class. Quant. Grav.,
{\bf 10}, 1825 (1993); Phys. Rev. D {\bf 51}, 2713 (1995).

\bibitem{cornish} N. J. Cornish and J. J. Levin, Phys. Rev. D {\bf
53}, 3022 (1996).

\bibitem{ivano} J. Guckenheimer and P. Holmes, {\it Dynamical
Systems and Bifurcations of Vector Fields}, Appl. Math. Sciences,
Vol. 42, Springer-Verlag, New York (1983); J. Koiller, J. R. T. M.
Neto and I. Dami\~ao Soares, Phys. Lett. A {\bf 110}, 260 (1985).

\bibitem{oss} H. P. de Oliveira, I. Dami\~ao Soares and T. J. Stuchi,
Phys. Rev. D {\bf 56}, 730 (1997); H. P. de Oliveira, I. Dami\~ao Soares and
T. J. Stuchi, {\it Chaotic exit to inflation: the dynamics of preinflationary
universes}, preprint gr-qc/9711014 and G. A. Monerat, H. P. de Oliveira and
I. Dami\~ao Soares, {\it Chaotic exits to inflation: preinflationary
Friedmann-Robertson-Walker universes}, preprint gr-qc/9711023, to appear in
the Proceedings of the VIII Marcel Grossman Meeting.

\bibitem{mtw} C. W. Misner, K. S. Thorne and J. A. Wheeler, {\it
Gravitation}, San Francisco: Freeman (1973).

\bibitem{hawking} J. Halliwell and S. Hawking, Phys. Rev. {\bf D}
31, 1777 (1985).

\bibitem{wiggins} S. Wiggins, {\it Global Bifurcations and Chaos},
Springer-Verlag, Berlin Heidelberg (1988).

\bibitem{Moser} M. A. Moser, Commun. Pure Appl. Math., {\bf 11}, 257 (1958).

\bibitem{ozorio} A. M. Os\'{o}rio de Almeida, N. De Leon, M. A.
Metha and C. C. Marston, Physica D {\bf 46}, 265 (1990).

\bibitem{vieira} W. M. Vieira and A. M. Os\'{o}rio de Almeida, Physica D {\bf
90}, 9 (1996).

\bibitem{berry} M. V. Berry, {\it Regular and Irregular Motion}, AIP
Conference Proccedings, n. 46, Editor: S. Jorna, American Physcis
Institute (1978).

\bibitem{ed} E. S. Cheb-Terrab and H. P. de Oliveira, Comp. Phys.
Commun. 95, 171 (1996).

\bibitem{os} Chaos and structure formation, H. P. de Oliveira and I. Dami\~ao
Soares, work in progress. 

\bibitem{linde} A. Linde, {\it Particle Physics and Inflationary Cosmology},
Harwood Academic Publishers (1993). 

\end{thebibliography}
\end{document}